\RequirePackage{fix-cm}
\documentclass[twocolumn,epjc3]{svjour3}
\smartqed  
\RequirePackage{graphicx}
\journalname{}
\begin{document}

\title{Analytical model of millisecond pulsar PSR J0514 - 4002A}


\author{Sajahan Molla\thanksref{e1,addr1}
        \and
        Bidisha Ghosh\thanksref{e2,addr2}
        \and
        Mehedi Kalam\thanksref{e3,addr2} 
}

\thankstext{e1}{e-mail: sajahan.phy@gmail.com}
\thankstext{e2}{e-mail: bidishaghosh.physics@gmail.com}
\thankstext{e3}{e-mail: kalam@associates.iucaa.in}

\institute{Department of Physics, New Alipore College, L Block, New Alipore, Kolkata 700053, India \label{addr1}
           \and
           Department of Physics, Aliah University, IIA/27, Action Area II, Newtown, Kolkata 700156, India \label{addr2}
}

\date{Received: date / Accepted: date}

\maketitle

\begin{abstract}
We construct a relativistic model for the newly discovered
millisecond pulsar PSR J0514 - 4002A located in the globular
cluster NGC 1851 (A. Ridolfi, P.C.C. Freire, Y. Gupta, S.M.
Ransom, MNRAS 490, 3860 (2019)) by using Tolman VII spacetime. We
have obtained central density ($\rho_{0}$), central pressure
($p_{0}$), probable radius, compactness ($u$) and surface
redshift ($Z_{s}$) of the above mentioned newly discovered
millisecond pulsar, which is very much consistent with reported
data. Equation of State (EoS) of the millisecond pulsar has comes
out as stiff in nature which is physically acceptable. Not even
that our proposed model can analyze most of the millisecond
pulsars having masses up to $1.51 M_{\odot}$.
\end{abstract}

\section{Introduction}
\label{intro} The low-mass X-ray binaries (LMXBs) are binary
systems in which a neutron star or a black hole accretes matter
from a low-mass companion star. These systems(LMXBs) provide a
laboratory to test the behavior of matter under extreme physical
conditions that are unavailable on Earth. The reprocessed X-ray
emission in the outer accretion disc, which is formed around the
compact object, dominates the optical flux thereby overflowing
any intrinsic spectral feature of the donor star
\cite{Cornelisse2013}. The Millisecond pulsars (MSPs) are thought
to be the offspring of low-mass X-ray binaries (LMXBs) in which
mass transfer from a low-mass companion star to the neutron star
carries angular momentum and spins it up to a very fast rotation
\cite{Alpar1982}. When the rate of mass transfer decreases in the
later evolutionary stages, these binaries host a radio
millisecond pulsar \cite{Backer1982,Ruderman1989}. At the end of
their spin-up phase they are reactivated as rotation-powered
(radio and $\gamma$-ray loud) pulsars. In the year 1993,
Rotation-powered MSPs were identified as pulsed X-ray sources
\cite{Becker1993}. The Millisecond pulsars (MSPs) are apparently
long aged and emit optical, X-ray and $\gamma$-ray fluxes
significantly below from the awaited canonical pulsars with
similar periods. The first MSP B1937+214, by the definition of
fast accretion spun-up pulsar was discovered using the Arecibo
telescope in 1982 \cite{Backer1982}. A number of recently
discovered millisecond pulsars (MSPs), which lie in binary
systems with evolved companion (neutron star), tend to be lighter
than the companion due to reduced mass transfer
\cite{Antoniadis2016}. The study of Millisecond pulsars (MSPs)
takes much attention to the astrophysicists in the present decades
\cite{Zilles2020,Voisin2020,Gusinskaia2020,Vivekanand2019,Bogdanov2019,Riley2019,Bult2019,Webb2019}.
This study is very important in the physical as well as in
theoretical point of view. Scientists used different techniques
(such as observational, computational or theoretical analysis) to
study the physical properties of astrophysical compact objects
like MSPs. The small size massive compact objects like white
dwarfs, neutron stars, pulsars, MSPs or black hole, has high dense
configurations which influence researchers to obtain models that
describe compact objects in the regime of strong gravitational
field.

General Relativity describes the gravitational interaction and
its consequences very well in a four-dimensional spacetime and it
has been confirmed by Will \cite{Will2005} with observational and
experimental support. The Einstein's theory of general relativity
laid the foundation of our understanding of compact stars. The
exact solution of Einstein's field equations was first solved by
Schwarzschild \cite{Schwarzschild1916} in 1916. He had calculated
the gravitational field of a homogeneous sphere of finite radius,
which consists of incompressible fluid. In 1939,Oppenheimer,
Volkoff and Tolman \cite{Oppenheimer1939,Tolman1939} successfully
derived the balancing equations of relativistic stellar
structures from Einstein's field equations and they discovered
the limit to the mass of a stable relativistic incompressible
fluid sphere. With the discovery of pulsars, it was the demand
from the scientific community to study neutron stars to build up
a connecting bridge between theoretical and observational
physics. Compact star models give interesting results in term of
the important parameters like stability factor, compactness,
red-shift, equation of state etc. which cannot be inferred from
direct observation. We suggest to see the details work carried
out by many researchers
\cite{Rahaman2012a,Rahaman2012b,Kalam2012a,Hossein2012,Kalam2012b,Kalam2013,Kalam2014,Kalam2014a,Hossein2016a,Hossein2016b,Kalam2016,Kalam2017,Kalam2018,Molla2019,Islam2019,Hendi2016,Hendi2017,Panah2017,Ngubelanga2015,Paul2015,Lobo2006,Bronnikov2006,Egeland2007,Dymnikova2002}
on compact stars.

Tolman \cite{Tolman1939} proposed an static solutions
for a fluid sphere. He pointed out that due to some complexity of
the VII-th solution (among the eight different solutions), it is
not possible to explain the physical behavior of a star. We have
some curiosities about these conclusion. We thought this solution
may explore some physics to compact stars. Several studies have
been done on compact stars by many researchers using the Tolman
VII-th solution till now. Among them, Neary, Ishak and Lake
\cite{Neary2001} have found that apart from the exact static
spherically symmetric perfect fluid solution, Tolman VII solution
exhibits a surprising good approximation for neutron star. The
Tolman VII solution is the first exact causal solution which has
shown to exhibits trapped null orbits for a tenuity (total radius
to mass ratio) $>3$. Thomas E. Kiess \cite{Kiess2012} derived an
exact Einstein-Maxwell metric for a static spherically symmetric
perfect fluid with mass and charge. Addition of modest charge to a
neutral star enables it to have a larger total mass, a different
radius and a larger red-shift. Raghoonundun and Hobill
\cite{Raghoonundun2015} had deduced for the first time a closed
form class of equation of states (EoSs) for neutron stars by
using Tolman VII solution. The EoSs allow further analysis of it,
leading to a viable model for compact stars with arbitrary surface
mass density. It also obeys the stability criteria. Bhar et al.
\cite{Piyali2015} have studied the behavior of static spherically
symmetric relativistic stellar objects with locally anisotropic
matter distribution considering the Tolman VII spacetime. They
have analyzed different physical properties of the stellar model
and presented it graphically. Singh et al. \cite{Singh2016} have
presented a new exact solution of charged anisotropic Tolman VII
type solution representing compact stars. According to them, this
solution can be used to model both neutron star and quark star
within the range of observed masses and radii. Bhar et al.
\cite{Piyali2017} also have studied compact star in Tolman VII
spacetime with quadratic equation of state of matter
distribution.Using their solution, they optimized the masses and
radii of few well-known compact stars within the limit of their
observed values. Sotani and Kokkotas \cite{Sotani2018} have
systematically examine the compactness of neutron stars using
Tolman VII solutions in scalar-tensor theory of gravity. They
showed that the maximum compactness of neutron stars in general
relativity is higher than that in scalar-tensor gravity when the
coupling constant is confined to values provided by astronomical
observations. Hensh and  Stuchlík \cite{Hensh2019} have started
from the isotropic Tolman VII perfect fluid solution and by using
the MGD method they have got a new exact and analytical solution.
This new solution represents the anisotropic version of the
Tolman VII solution, which satisfies all criteria for physical
acceptability. Hence, they argued that this new solution could be
used to model of stellar configurations, like neutron stars. They
demonstrate that the effect of anisotropy will increase with the
increases in coupling constant($\alpha$).

Ridolfi et al. \cite{Ridolfi2019} very recently reported the
results of one year of upgraded Giant Metrewave Radio Telescope
timing measurements of PSR J0514-4002A, a 4.99-ms pulsar in a
18.8-day, eccentric (e = 0.89) orbit with a massive companion
located in the globular cluster NGC 1851. They greatly improve
the precision of the rate of advance of periastron,
$\dot{\omega}$ = 0.0129592(16) deg $yr^{-1}$ by combining these
timing measurements data with the earlier Green Bank Telescope
data. As a result, a much refine measurement of the total mass
$M_{tot} =2.4730(6)M_{\odot}$ of the binary has been done. They
also measured the Einstein delay parameter,$\gamma$ = 0.0216(9)s
for this binary pulsar which has never been done for any binary
system with an orbital period larger than $\sim$ 10 h. Besides,
they measured the proper motion of the system ($\mu_{\alpha} =
5.19(22)$ \& $\mu_{\delta} = -0.56(25)$ mas $ yr^{-1}$), which is
not only important for analyzing its motion in the cluster, but
also essential for a proper interpretation of $\gamma$. They
obtained one of the lowest ever measured a pulsar mass of $M_{p}=
1.25^{+0.05}_{-0.06}M_{\odot}$ and a companion mass of $M_{c}
=1.22^{+0.06}_{-0.05}M_{\odot}$. They guessed that the companion
may be a neutron star.

In 2016, Reardon et al. \cite{Reardon2016} has measured the mass
of PSR J0437-4715 as $1.44 \pm 0.07 M_{\odot}$. After
approximately 2 years of observations with the
Caltech-Parkes-Swinburne Recorder II, Jacoby et al.
\cite{Jacoby2005} precisely measured millisecond pulsar PSR
J1909-3744 mass as $1.438 \pm 0.024 M_{\odot}$. Nice et al. in
2008 \cite{Nice2008} and Fortin et al. in 2016 \cite{Fortin2016}
has also reported the mass of millisecond pulsar PSR J0751+1807 as
$1.26 \pm 0.14 M_{\odot}$.

In this paper, we want to study the physical behaviour of the
newly discovered millisecond pulsar PSR J0514-4002A. For this we
have considered the anisotropic model to
study the fluid sphere. Our main objective in this study is to
give an estimate of the equation of state(EoS) of nuclear matter
as well as the possible radius of the newly discovered millisecond
pulsar PSR J0514-4002A.

We organize the article as follows: In Sec. 2, we have discussed
the interior spacetime; density, pressure behavior and EoS of the
millisecond pulsar. In Sec. 3, we have studied some special
features like: Exterior spacetime and matching condition,
Generalized TOV equation, Energy conditions, Stability,
Compactness and Surface red-shift in different sub-sections. In
Sec. 4, we have concluded our discussion with numerical data.

\section{Interior spacetime and Equation of State}
\label{sec:1}  Being inspired from many previously revealed
famous articles
\cite{Takisa2019,Matondo2018,Maurya2015,Maurya2019,Dayanandana2017,Maurya2017,Gedela2018,Singh2017,Jafry2017,Rahaman2017,Jasim2016,Takisa2016,Maurya2016,Singh2017a},
we have taken static, spherical symmetric metric for this pulsar
 as:
\begin{equation}
ds^2=-e^{\nu(r)}dt^2 + e^{\lambda(r)}dr^2 +r^2(d\theta^2 +sin^2\theta d\phi^2) \label{eq1}
\end{equation}
where $\lambda$ and $\nu$ are the functions of radial
coordinate(r).\\
For our model, the form of the energy-momentum tensor is
\begin{equation}
               T_\nu^\mu=  ( -\rho , p_{r}, p_{t}, p_{t})  \label{Eq2}
\end{equation}
where $\rho$ is the energy density, $p_{r}$ $\&$ $p_{t}$ are the radial and transverse pressure respectively.\\

Einstein's field equations for the metric (1) accordingly are
obtained as ($ c=1,G=1$)
\begin{eqnarray}
8\pi  \rho &=& e^{-\lambda}\left[\frac{\lambda^\prime}{r}-\frac{1}{r^2}\right]+\frac{1}{r^2},\label{eq3}\\
8\pi  p_{r} &=& e^{-\lambda}\left[\frac{\nu^\prime}{r}+\frac{1}{r^2}\right]-\frac{1}{r^2}.\label{eq4} \\
8\pi  p_{t} &=& \frac{e^{-\lambda}}{2}\left[\nu^{\prime\prime}+\frac{\nu^\prime-\lambda^\prime}{r}+\frac{{{\nu^\prime}^2}-{\lambda^\prime\nu^\prime}}{2}\right] \label{eq5}
\end{eqnarray}

To solve above Einstein's field equations we have considered
Tolman VII solution as \cite{Tolman1939}
\begin{eqnarray}
ds^2 = -B^2\sin ^{2}\ln \sqrt{\frac{\sqrt{1-\frac{r^2}{R^2}+4\frac{r^4}{A^4}}+2\frac{r^2}{A^2}-\frac{1}{4}\frac{A^2}{R^2}}{C}}dt^2 \nonumber \\
+\left(1-\frac{r^2}{R^2}+4\frac{r^4}{A^4}\right)^{-1}dr^2 +r^2d\Omega^{2}. ~~~~~~~~~~\label{eq61}
\end{eqnarray}
where $R(length)$, $C(dimensionless)$, $A(length)$ and $B$ $(length/time)$ are constants which will be determined later using some physical boundary conditions.\\
Now from equation (6) we get,
\begin{eqnarray}
e^{-\lambda} &=& \left(1-\frac{r^2}{R^2}+4\frac{r^4}{A^4}\right) \label{eq7}
\end{eqnarray}
\begin{eqnarray}
\lambda^\prime &=&
\frac{\left(2\frac{r}{R^2}-16\frac{r^3}{A^4}\right)}{\left(1-\frac{r^2}{R^2}+4\frac{r^4}{A^4}\right)}, \label{eq8}
\end{eqnarray}
\begin{eqnarray}
\nu^\prime &=& \frac{\left(1-\frac{r^2}{R^2}+4\frac{r^4}{A^4}\right)^{-1/2}\left(-\frac{r}{2R^2}+4\frac{r^3}{A^4}\right)+2\frac{r}{A^2}}
{\left(\sqrt{1-\frac{r^2}{R^2}+4\frac{r^4}{A^4}}+2\frac{r^2}{A^2}-\frac{1}{4}\frac{A^2}{R^2}\right)}\nonumber\\
&&2\cot\ln\sqrt{\frac{\sqrt{1-\frac{r^2}{R^2}+4\frac{r^4}{A^4}}+2\frac{r^2}{A^2}-\frac{1}{4}\frac{A^2}{R^2}}{C}},\label{eq9}
\end{eqnarray}

\begin{eqnarray}
\nu^{\prime\prime} &=& [\left(\sqrt{1-\frac{r^2}{R^2}+4\frac{r^4}{A^4}}+2\frac{r^2}{A^2}-\frac{1}{4}\frac{A^2}{R^2}\right)^{-1} \nonumber
\\&& \times \{-2\left(1-\frac{r^2}{R^2}+4\frac{r^4}{A^4}\right)^{-3/2}\left(-\frac{r}{2R^2}+4\frac{r^3}{A^4}\right)^{2} \nonumber
\\&& +\left(1-\frac{r^2}{R^2}+4\frac{r^4}{A^4}\right)^{-1/2}\left(-\frac{1}{2R^2}+\frac{12r^2}{A^4}\right)+\frac{2}{A^2}\} \nonumber
\\&& -2\left( \sqrt{1-\frac{r^2}{R^2}+4\frac{r^4}{A^4}}+2\frac{r^2}{A^2}-\frac{1}{4}\frac{A^2}{R^2} \right)^{-2} \nonumber
\\&& \times \{ \left(1-\frac{r^2}{R^2}+4\frac{r^4}{A^4}\right)^{-1/2}\left(-\frac{r}{2R^2}+4\frac{r^3}{A^4}\right)+\frac{2r}{A^{2}} \}^{2}] \nonumber
\\&& \times 2\cot\ln\sqrt{\frac{\sqrt{1-\frac{r^2}{R^2}+4\frac{r^4}{A^4}}+2\frac{r^2}{A^2}-\frac{1}{4}\frac{A^2}{R^2}}{C}} \nonumber
\\&& -2[\frac{\left(1-\frac{r^2}{R^2}+4\frac{r^4}{A^4}\right)^{-1/2}\left(-\frac{r}{2R^2}+4\frac{r^3}{A^4}\right)+\frac{2r}{A^{2}}}
{\sqrt{1-\frac{r^2}{R^2}+4\frac{r^4}{A^4}}+2\frac{r^2}{A^2}-\frac{1}{4}\frac{A^2}{R^2}} \nonumber
\\&& \times \mathrm{cosec}\ln\sqrt{\frac{\sqrt{1-\frac{r^2}{R^2}+4\frac{r^4}{A^4}}+2\frac{r^2}{A^2}-\frac{1}{4}\frac{A^2}{R^2}}{C}}]^{2}
\label{eq10}
\end{eqnarray}

$\bf{Solution~of~Density~and~Pressure:}$\\

Now from eqn.(3), eqn.(7) and eqn.(8) we get,
\begin{eqnarray}
\rho &=& \frac{1}{8\pi}\left(\frac{3}{R^2}-20\frac{r^2}{A^4}\right)
\end{eqnarray}
\begin{eqnarray}
\frac{d\rho}{dr} &=& - \frac{5r}{\pi  A^4}< 0,\nonumber\\
\frac{d\rho}{dr} (r=0) &=& 0,\nonumber \\
\frac{d^2 \rho}{dr^2}(r=0) &=& -\frac{5}{\pi  A^4} <0.\nonumber
\end{eqnarray}

We also get from eqn.(4), eqn.(7) and eqn.(9),
\begin{eqnarray}
p_{r} &=& \frac{\left(1-\frac{r^2}{R^2}+4\frac{r^4}{A^4}\right)}{8\pi}\nonumber\\
&&\times[\frac{\left(1-\frac{r^2}{R^2}+4\frac{r^4}{A^4}\right)^{-1/2}\left(4\frac{r^2}{A^4}-\frac{1}{2R^2}\right)+\frac{2}{A^2}}
{\left(\sqrt{1-\frac{r^2}{R^2}+4\frac{r^4}{A^4}}+2\frac{r^2}{A^2}-\frac{1}{4}\frac{A^2}{R^2}\right)}\nonumber\\
&&2\cot\ln\sqrt{\frac{\sqrt{1-\frac{r^2}{R^2}+4\frac{r^4}{A^4}}+2\frac{r^2}{A^2}-\frac{1}{4}\frac{A^2}{R^2}}{C}}+\frac{1}{r^2}]\nonumber\\
&&-\frac{1}{8\pi r^2} \label{eq4}
\end{eqnarray}
\begin{eqnarray}
\frac{dp_{r}}{dr}(r=0) &=& 0\nonumber\\
\frac{d^2 p_{r}}{dr^2}(r=0) &=&  < 0\nonumber
\end{eqnarray}
 Variations of the energy-density, radial and transverse
pressure in the stellar interior of the millisecond pulsar PSR
J0514-4002A are shown in Fig.~1, Fig.~2 and Fig.~3 respectively.
From the figure we observe that
 $\rho(r)$, $p_{r}(r)$ and $p_{t}(r)$ are monotonically decreasing function of $r$ and
they are maximum at the center of the star. Therefore, the energy
density and the anisotropic pressure are well behaved in the
interior of the stellar structure. The anisotropic parameter
$\Delta = \left( p_t - p_r \right)$ representing the anisotropic
stress in the stellar interior is shown in Fig.~4.
\begin{figure}
\includegraphics[width=0.50\textwidth]{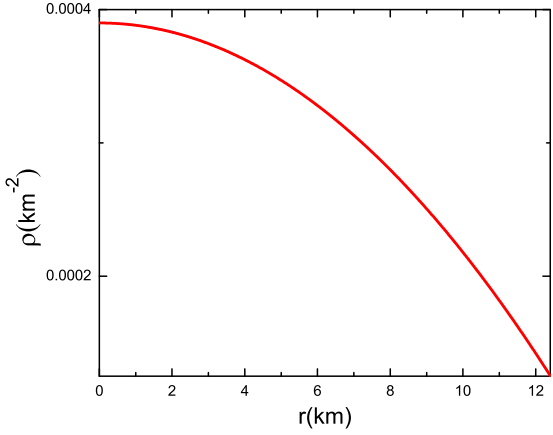}
\caption{Variation of the energy-density, $\rho(r)$ within the
stellar interior of the millisecond pulsar PSR J0514-4002A.}
\end{figure}
\begin{figure}
\includegraphics[width=0.50\textwidth]{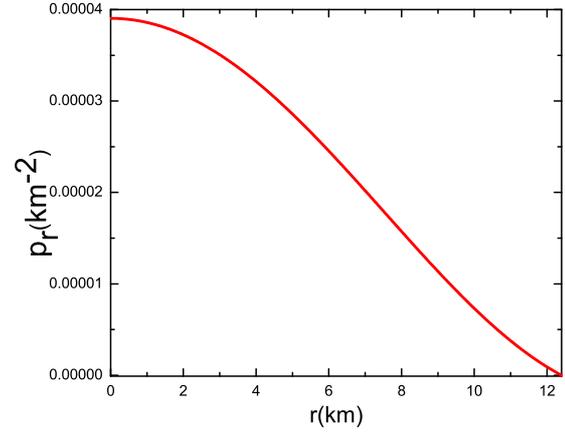}
\caption{Variation of the radial pressure $p_{r}(r)$ within the
stellar interior of the millisecond pulsar PSR J0514-4002A.}
\end{figure}
\begin{figure}
\includegraphics[width=0.50\textwidth]{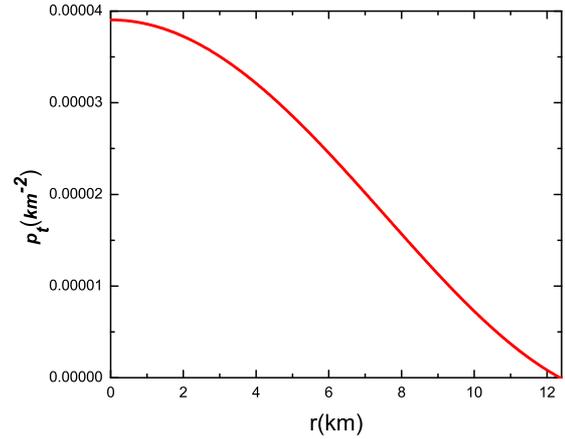}
\caption{Variation of the transverse pressure $p_{t}(r)$ within
the stellar interior of the millisecond pulsar PSR J0514-4002A.}
\end{figure}
\begin{figure}
\includegraphics[width=0.50\textwidth]{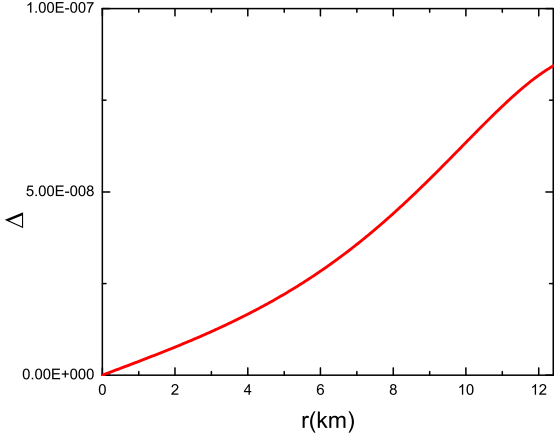}
\caption{Variation of the anisotropic parameter, $\Delta$ within
the stellar interior of the millisecond pulsar PSR J0514-4002A.}
\end{figure}

$\bf{Equation~of~State:}$ \\

\label{sec:2} We know that different Equation of State (EoS) leads
to different matter distribution of the pulsar and hence different
Mass-Radius relation. Here, we have obtained a relation between
density $(\rho)$ and radial pressure ($p_{r}$) which is known as Equation of
State (EoS) by using curve fitting technique and interestingly it
comes out to be polytropic in nature. The relation trace out
given below:
\begin{eqnarray}
 p_{r} = \alpha \rho^{\beta}
\end{eqnarray}
where $\alpha$ is the polytropic constant and $ \beta$ is related
to polytropic index.  As we know that $ \beta = 2$ implies for
nonrelativistic degenerate fermions, whereas $ \beta = 1$ leads to
stiff matter EoS. In our model, the numerical value of $\alpha $
and $\beta $ are comes out as $\alpha$=2.462$\times 10^{-7}$ and
$ \beta$=2.856 which indicates that the equation of state (EoS)
of the pulsar is stiff(Fig.~5) \cite{ozel2006,Guo2014,Lai2009}.
\begin{figure}
\centering
\includegraphics[width=0.50\textwidth]{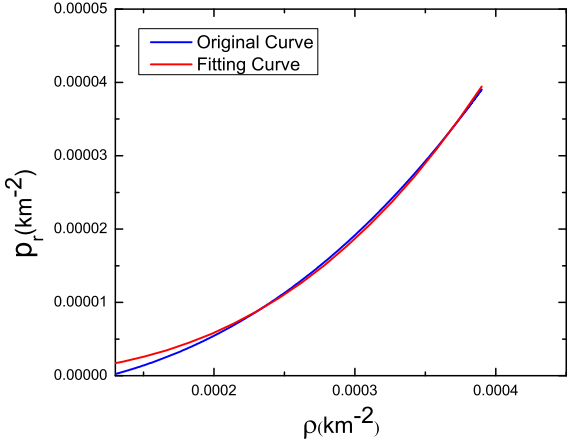}
\caption{The equation of state (EoS) within the stellar interior
of the millisecond pulsar PSR J0514-4002A.}
\end{figure}

\section{Physical analysis}

\subsection{Exterior spacetime and matching condition}
Our interior solution should match the exterior
Schwarzschild
metric at the boundary ($r=b$) where $b$ is the radius of the
star. The exterior spacetime is given by the Schwarzschild line
element
\begin{eqnarray}
ds^2 = - \left(1-\frac{2M}{r}\right)dt^2 +  \left(1-\frac{2M}{r}\right)^{-1}dr^2 \nonumber\\
+r^2(d\theta^2 +sin^2\theta d\phi^2) \label{eq1}
\end{eqnarray}
Now at the boundary $r=b$ the coefficients of $g_{tt}$, $g_{rr} $ and $\frac{\partial g_{tt}}{\partial r}$ all are continuous. This implies
\begin{eqnarray}
\left(1-\frac{b^2}{R^2}+4\frac{b^4}{A^4}\right) &=& 1 - \frac{2M}{b},\label{eq13}
\end{eqnarray}
\begin{eqnarray}
B^2\sin ^{2}\ln \sqrt{\frac{\sqrt{1-\frac{b^2}{R^2}+4\frac{b^4}{A^4}}+2\frac{b^2}{A^2}-\frac{1}{4}\frac{A^2}{R^2}}{C}}  \nonumber\\
 = \left(1-\frac{2M}{b}\right) \label{eq14}
\end{eqnarray}
 and
\begin{eqnarray}
\left(-\frac{2b}{R^2}+16\frac{b^3}{A^4}\right) &=& \frac{2M}{b^2} \nonumber
\end{eqnarray}
Solving above equations we get,
\begin{eqnarray}
R &=& \frac{b^{3/2}}{\sqrt{5M}} \nonumber
\end{eqnarray}
\begin{eqnarray}
A &=& \frac{\sqrt{2} b^{5/4}}{(3M)^{1/4}} \nonumber
\end{eqnarray}
Furthermore, using the expression (12) for $p_{r}$(r=b)=0 we obtain
\begin{eqnarray}
C &=& \frac{\sqrt{1-\frac{b^2}{R^2}+4\frac{b^4}{A^4}}+2\frac{b^2}{A^2}-\frac{1}{4}\frac{A^2}{R^2}}{e^{\delta}} \nonumber
\end{eqnarray}
Where,
\begin{eqnarray}
\delta &=& \left( 2\cot^{-1} \{ \frac{\frac{1}{b^2\left(1-\frac{b^2}{R^2}+4\frac{b^4}{A^4}\right)}-\frac{1}{b^2}}{2 \frac{\left(1-\frac{b^2}{R^2}+4\frac{b^4}{A^4}\right)^{-1/2}\left(4\frac{b^2}{A^4}-\frac{1}{2R^2}\right)+\frac{2}{A^2}}
{\left(\sqrt{1-\frac{b^2}{R^2}+4\frac{b^4}{A^4}}+2\frac{b^2}{A^2}-\frac{1}{4}\frac{A^2}{R^2}\right)}} \} \right) \nonumber
\end{eqnarray}

Now from the equation ~(\ref{eq13}), we get the compactification factor as
\begin{equation}
u = \frac{M}{b} = \left(\frac{b^2}{2R^2}-2\frac{b^4}{A^4}\right).\label{eq15}
\end{equation}

\subsection{Equilibrium analysis via generalized TOV-equation}
 To check whether our model is in equilibrium under three
different forces($F_g$, $F_h$ and  $F_a$), we consider the
generalized Tolman-Oppenheimer-Volkov (TOV) equation which is
represented by the formula
\begin{equation}
\frac{dp_{r}}{dr} +\frac{1}{2} \nu^\prime\left(\rho+ p_{r}\right)+ \frac{2}{r}\left(p_{r}- p_{t}\right)= 0.\label{eq18}
\end{equation}
\begin{equation}
F_h + F_g + F_a  = 0,\label{eq21}
\end{equation}
where,
\begin{eqnarray}
F_h &=& -\frac{dp_r}{dr} \label{eq23} \\
F_g &=& -\frac{1}{2} \nu^\prime\left(\rho+p_r\right) \\
F_a &=& -\frac{2}{r}\left(p_{r}- p_{t}\right) \label{eq23}
\end{eqnarray}
Generalized TOV(stellar) equation describes the system is in
static equilibrium under gravitational ($F_g$), hydrostatic
($F_h$) and anisotropic ($F_a$) forces of the millisecond pulsar
PSR J0514-4002A. The profiles of the above mention three forces
of the pulsar are shown in Fig.~6. This figure shows that the
gravitational force is counterbalanced by the hydrostatic force
and anisotropic force, therefore consequently the present system
hold in static equilibrium.
\begin{figure}
\includegraphics[width=0.50\textwidth]{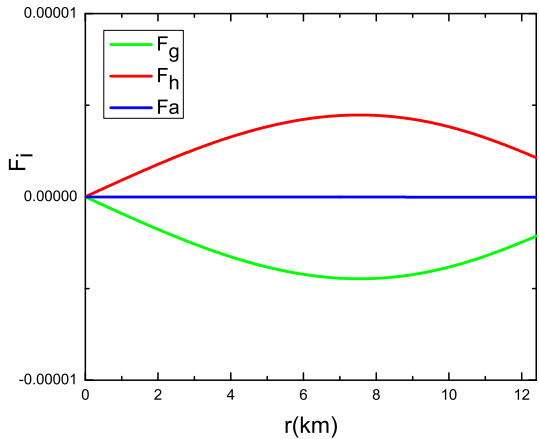}
\caption{The behavior of gravitational ($F_g$), hydrostatic
($F_h$) and anisotropic ($F_a$) forces at the stellar interior of
the millisecond pulsar PSR J0514-4002A.}
\end{figure}

\subsection{Energy conditions verification}
From the Fig.~7 we see that the following energy conditions like,
null energy condition (NEC), weak energy condition (WEC), strong
energy condition (SEC) and dominant energy
condition (DEC) are satisfied in our model :\\
(i) NEC: $p_r+\rho \geq0$ ,\\
(ii) WEC: $p_r+\rho \geq0$  , $~~\rho \geq0$  ,\\
(iii) SEC: $p_r+\rho \geq0$  ,$~~~~3p_r+\rho \geq0$ ,\\
(iv) DEC: $\rho > p_r $.
\begin{figure}
\includegraphics[width=0.50\textwidth]{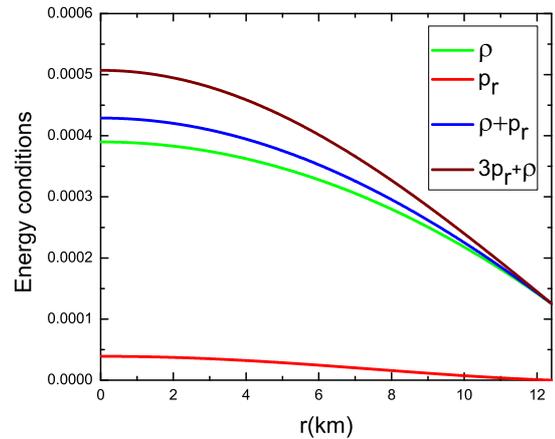}
\caption{Energy conditions at the stellar interior of the
millisecond pulsar PSR J0514-4002A.}
\end{figure}

\subsection{Stability checking}
Our proposed model for millisecond pulsar PSR J0514-4002A will be
physically acceptable if the sound speed within the stellar
interior is less than the speed of light (causality conditions),
where the sound speed can be obtained as
$v^2=\frac{dp}{d\rho}$\cite{Herrera1992,Abreu2007}.  In our
anisotropic pulsar model, we desire $v^2$ to be $0 \leq
v^2=(\frac{dp}{d\rho})\leq 1$. We plot the radial and transverse
sound speed for the millisecond pulsar PSR J0514-4002A in Fig.~8
\& Fig.~9 and observed that it satisfies well with the
inequalities $0\leq v_r^2 \leq 1$ and $0\leq v_t^2 \leq 1$
everywhere within the pulsar.  Therefore our pulsar model is well
stable.
\begin{figure}
\includegraphics[width=0.50\textwidth]{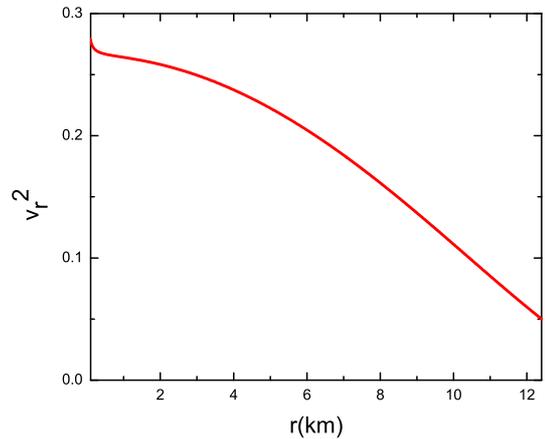}
\caption{Variation of the radial sound speed ($v_r^2$) at the stellar interior of the millisecond pulsar PSR J0514-4002A.}
\end{figure}
\begin{figure}
\includegraphics[width=0.50\textwidth]{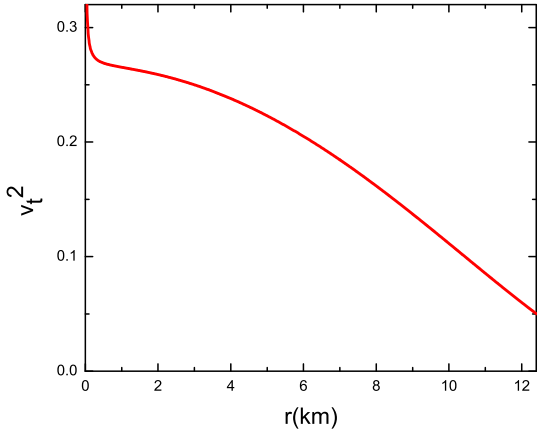}
\caption{Variation of the transverse sound speed ($v_t^2$) at the stellar interior of the millisecond pulsar PSR J0514-4002A.}
\end{figure}

We also verify the dynamical stability in presence of thermal
radiation. The adiabatic index($\gamma$) satisfies the condition
$\gamma = \frac{\rho+p_r}{p_r} \frac{dp_r}{d\rho} > \frac{4}{3}$
everywhere within the pulsar (Fig.~10). This type of stability
checking has previously done by several researcher
\cite{Chandrasekhar1964,Bondi1964,Bardeen1966,Knutsen1988,Harko2013}.
\begin{figure}
\includegraphics[width=0.50\textwidth]{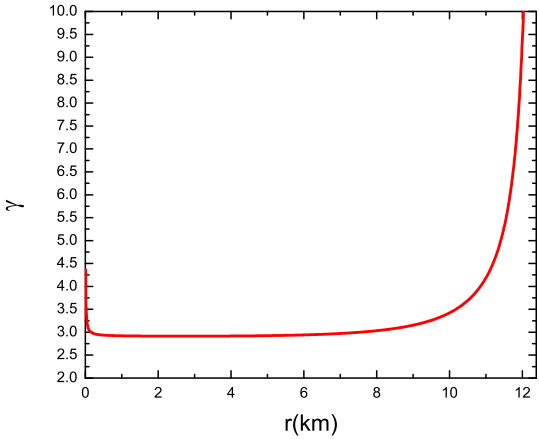}
\caption{Adiabatic index ($\gamma$) at the stellar interior of the
millisecond pulsar PSR J0514-4002A.}
\end{figure}

\subsection{Compactness and Surface redshift}
The mass function can be determined using the equation given below
\begin{equation}
\label{eq34}
 M(r)=4\pi\int^{b}_{0} \rho~~ r^2 dr =
 \frac{b}{2}\left[\frac{b^2}{R^2}-4\frac{b^4}{A^4}\right]
\end{equation}
where $b$ is the radius of the millisecond pulsar. According to
Buchdahl \cite{Buchdahl1959}, for a spherically symmetric perfect
fluid sphere, the allowable mass-radius ratio should
be $\frac{ Mass}{Radius} < \frac{4}{9}$.\\

The compactness, u is given by
\begin{equation}
\label{eq35} u= \frac{ M(b)} {b}=
 \frac{1}{2}\left[\frac{b^2}{R^2}-4\frac{b^4}{A^4}\right]
\end{equation}
The surface red-shift ($Z_s$) corresponding to the above
compactness ($u$) is
\begin{equation}
\label{eq36} 1+Z_s= \left[ 1-(2 u )\right]^{-\frac{1}{2}} ,
\end{equation}
where
\begin{equation}
\label{eq37} Z_s= \frac{1}{\sqrt{1-\frac{b^2}{R^2}+4\frac{b^4}{A^4}}}-1
\end{equation}

\begin{figure}
\includegraphics[width=0.50\textwidth]{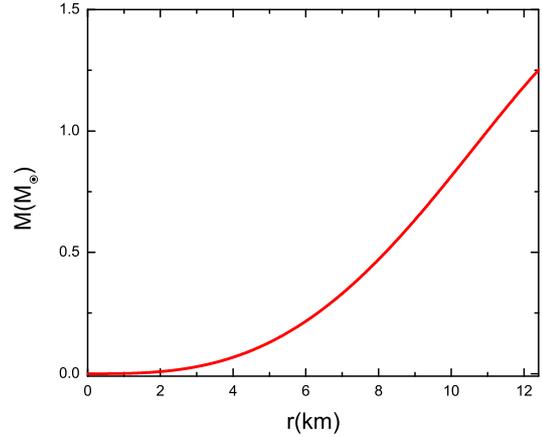}
\caption{Mass function, M(r) at the stellar interior of the
millisecond pulsar PSR J0514-4002A.}
\end{figure}

\begin{figure}
\includegraphics[width=0.50\textwidth]{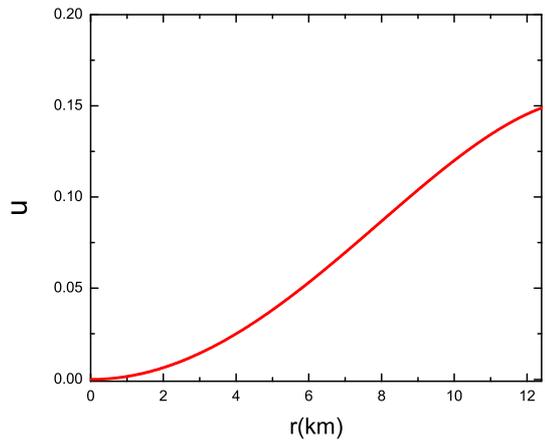}
\caption{Compactness (u) at the stellar interior of the
millisecond pulsar PSR J0514-4002A.}
\end{figure}

\begin{figure}
\includegraphics[width=0.50\textwidth]{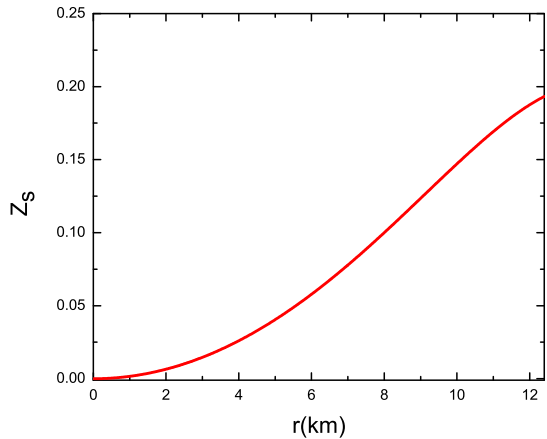}
\caption{Surface red-shift ($Z_{s}$) against radial parameter r
of the millisecond pulsar PSR J0514-4002A.}
\end{figure}

The variation of the mass function, compactness and surface
red-shift of the millisecond pulsar PSR J0514-4002A are shown in
Fig.~11, Fig.~12 and Fig.~13 respectively.

\begin{figure}
\centering
\includegraphics[width=0.50\textwidth]{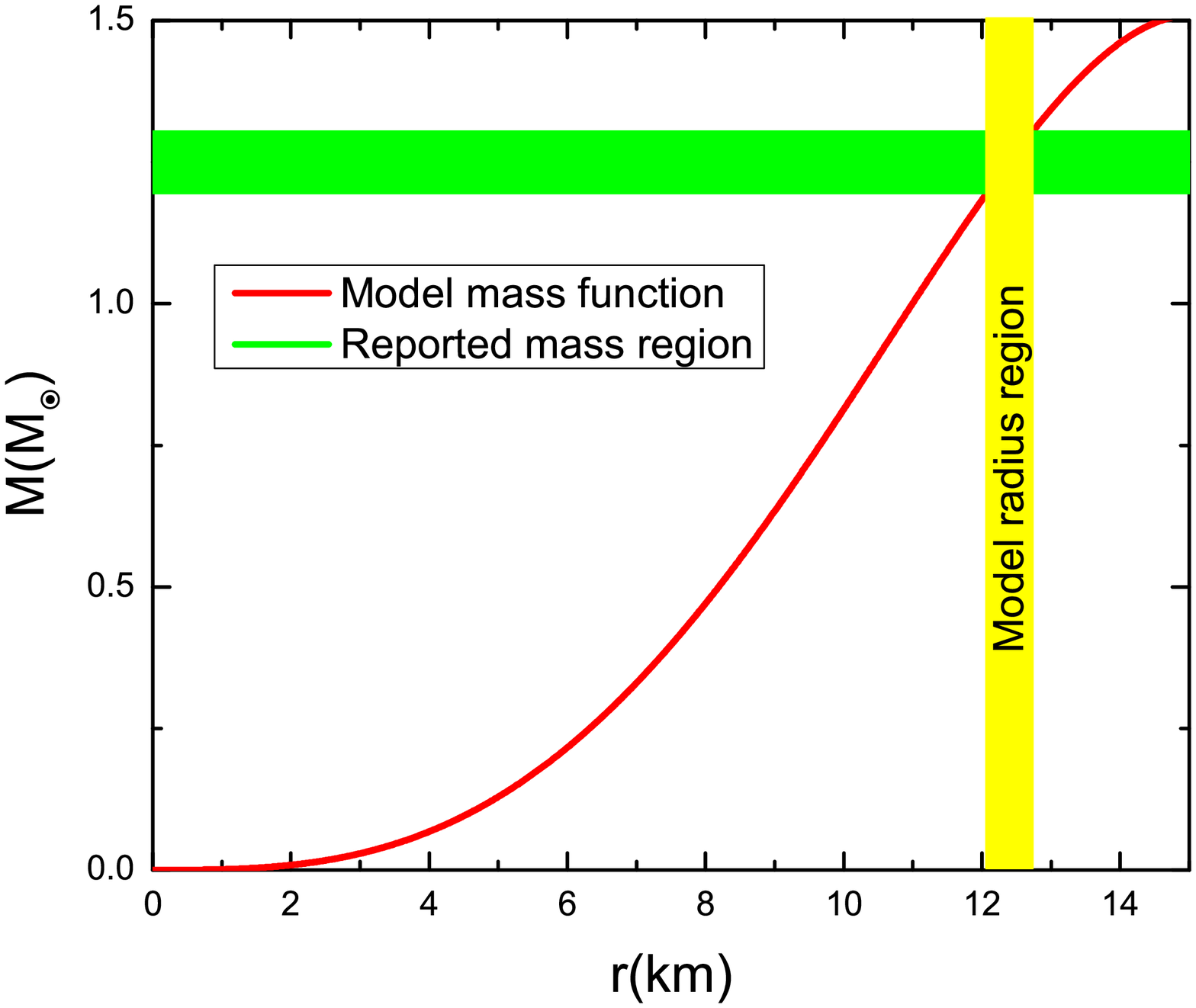}
\caption{Possible radius of the MSP PSR J0514-4002A.}
\end{figure}

\begin{figure}
\centering
\includegraphics[width=0.50\textwidth]{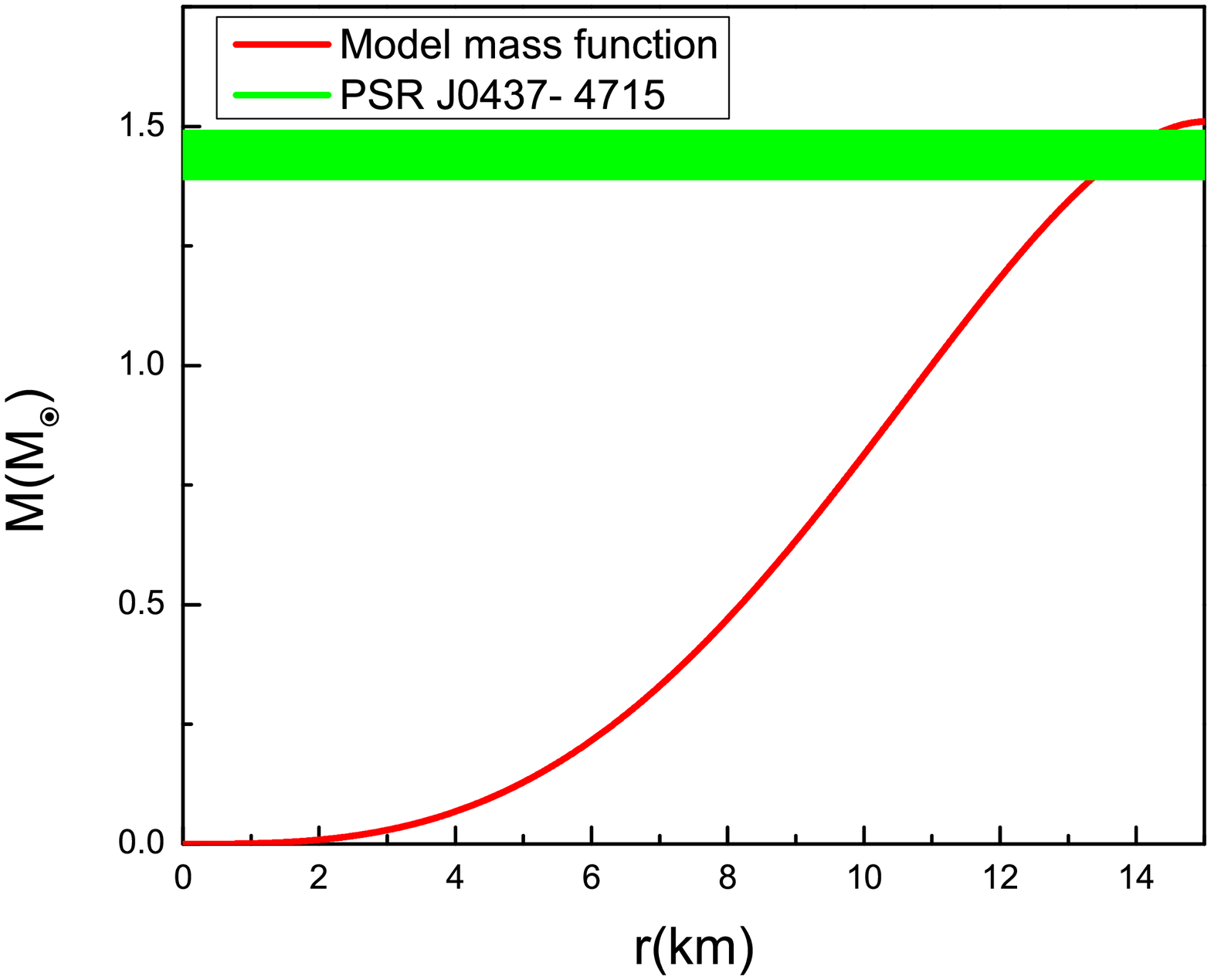}\\
\includegraphics[width=0.50\textwidth]{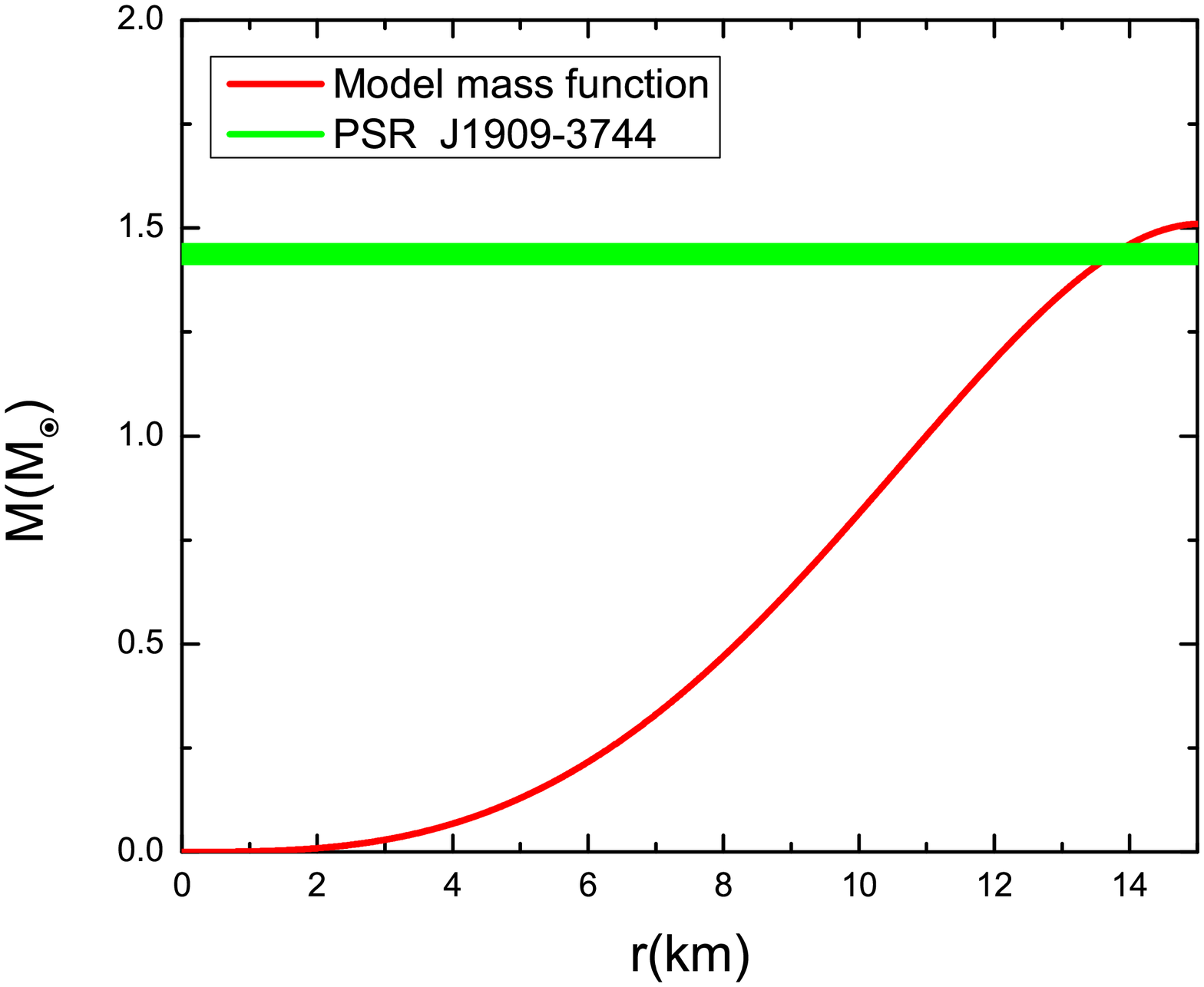}\\
\includegraphics[width=0.50\textwidth]{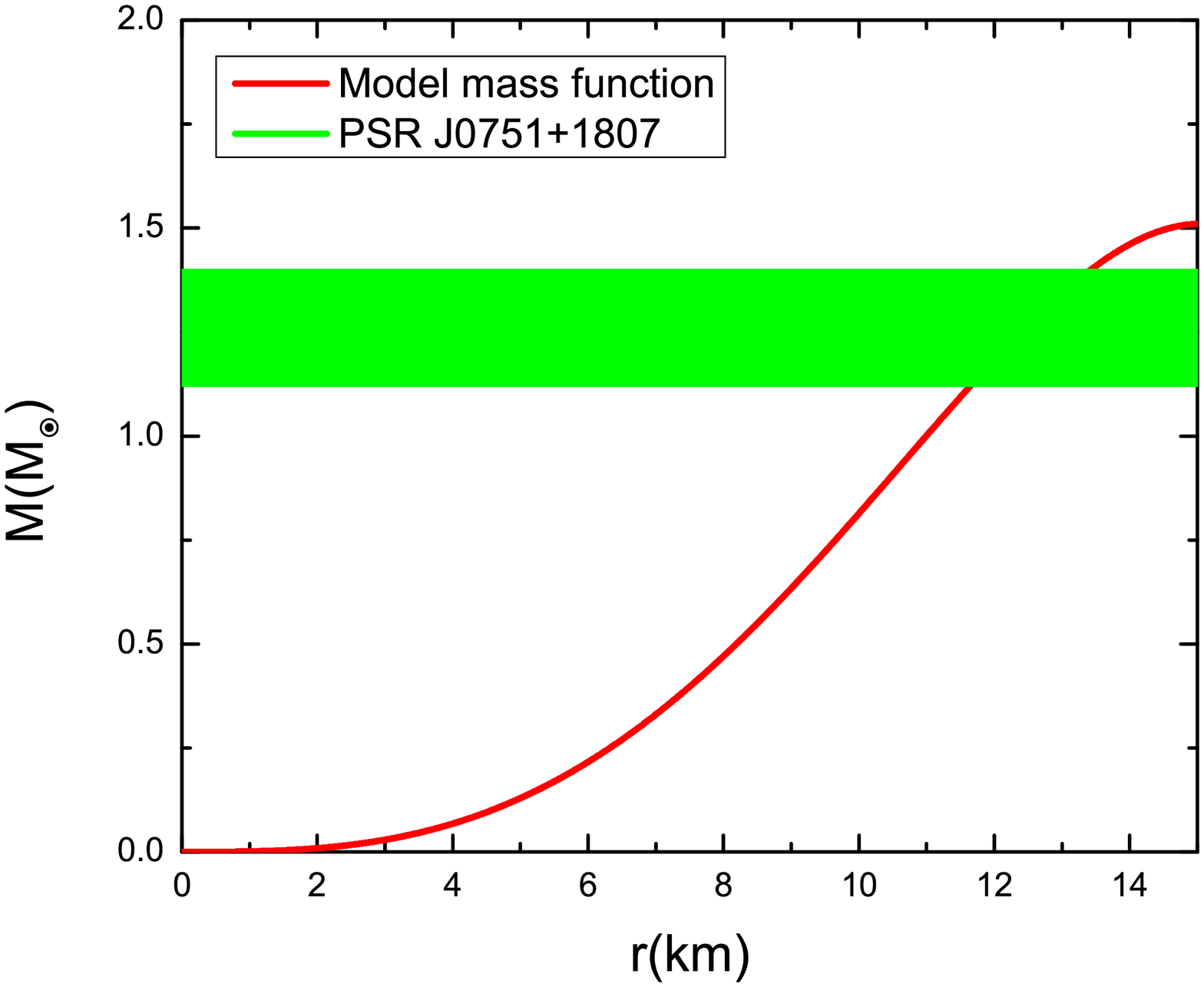}
\caption{Application of our model for other millisecond Pulsars.}
\end{figure}

\section{Discussion and concluding remarks}
In this paper, we have investigated the nature of the newly
discovered millisecond pulsar PSR J0514-4002A \cite{Ridolfi2019}
by considering Tolman VII \cite{Tolman1939} metric and taking
constituent matter as anisotropic in nature. Density and pressure of
the newly discovered millisecond pulsar PSR J0514-4002A are well
behaved (Figs. 1, 2, 3 and 4). Here, we have obtained a relation between
density $(\rho)$ and radial pressure ($p_r$) which is known as Equation of
State (EoS) by using curve fitting technique and interestingly it
comes out as polytropic in nature. The relation trace out as
$ p_r = \alpha \rho^{\beta}$
where $\alpha$ is the polytropic constant and $ \beta$ is related
to polytropic index. The numerical value of $\alpha $ and $\beta $
are comes out as $\alpha$=2.462$\times 10^{-7}$ and $
\beta$=2.856 which indicates that the equation of state (EoS) of
the pulsar is stiff(Fig.~5) \cite{ozel2006,Guo2014,Lai2009}.

Our interior solutions has been matched to the exterior
Schwarzschild line element at the boundary. Our model satisfies
stellar equation (generalised TOV) and all energy conditions (Fig
6 and 7). It is stable according to Herrera stability condition
\cite{Herrera1992} (Fig.~8 and 9). We have also check the dynamical
stability of the model in presence of thermal radiation
(Fig.~10). The mass function (M), compactness (u) and surface
red-shift ($Z_s$) has been plotted in Fig.~11, Fig.~12 and Fig.~13
respectively. The constants R, A, C can be determined in terms of
mass(M) and radius(b) by using the required physical boundary
conditions. Taking into account, the mass $M_{p}=
1.25^{+0.05}_{-0.06}M_{\odot}$ of the millisecond pulsar PSR
J0514-4002A obtained by Ridolfi et al. \cite{Ridolfi2019}, we
consider the values of the constants R = 17.5 km, A = 26.08 km and
C = 0.06713, such that the pressure drops from its maximum value
(at center) to zero at the boundary(Radius(b)= 12.39 km). As per
our model, the possible radius of the millisecond pulsar PSR
J0514-4002A has been found to be $12.39^{+0.31}_{-0.35} km$
(Fig.~14).
\begin{figure}
\includegraphics[width=0.50\textwidth]{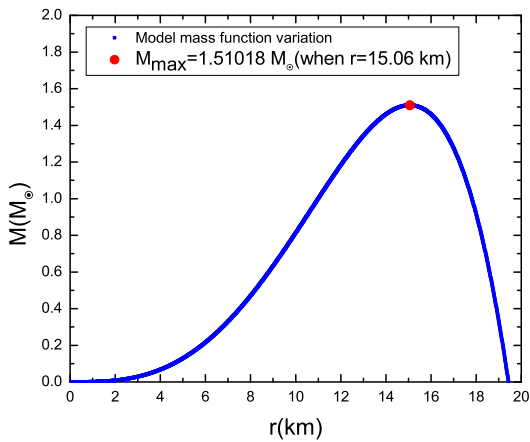}
\caption{Mass function, M(r) at the stellar interior
 with a maximum mass $1.51 M_{\odot}$.}
\end{figure}
The compactness ($u$) of the millisecond pulsar PSR J0514-4002A
has been found to be 0.14875 which satisfies Buchdahl limit
\cite{Buchdahl1959}. The surface red-shift ($Z_s$) for the
millisecond pulsar PSR J0514-4002A has also found to be 0.193
which is $\leq$ 0.8, implies the validity of our model
\cite{Haensel2000,Barraco2002}. It is to be mentioned here that
while solving Einstein's field equations, we set $c = G = 1$.
Now, inserting $G$ and $c$ into the relevant equations, the
values of the central density and central pressure of our pulsar
turn out to be $\rho_0$=0.52592$\times 10^{15}gm/cc$ (i.e.
0.000389767$km^{-2}$) and $p_0$=1.093$\times 10^{35}dyne/cm^2$
(i.e. 0.0000437354$km^{-2}$). It is worthy to be mention here that
our pulsar model is well applicable to other millisecond Pulsars
such as PSR J0437-4715, PSR J1909-3744, PSR J0751+1807 (Fig. 15).
From Fig. 15, one can easily estimate the probable radii of these
MSPs, once the radius is known with mass, other physical
parameters can be calculated. More interestingly, our model can
describe the characteristics of most of the pulsars having masses
upto $1.51 M_{\odot}$(Fig. 16).

Finally, we conclude by pointing out that our proposed model
(having stiff EoS) not only can be used to study the physical
properties of the millisecond pulsar PSR J0514-4002A, but also
for other millisecond pulsars (like: PSR J0437-4715, PSR
J1909-3744, PSR J0751+1807 etc.) having masses up to $1.51
M_{\odot}$.

\begin{acknowledgements}
MK like to thank IUCAA, Pune, India for providing research
facilities under Visiting Associateship where a part of this work
was carried out. BG  acknowledges Aliah University authority for
providing research fellowship under Ph.D. programme. We are
thankful to the respected referee for useful comments which help
us to improve the quality of the manuscript.
\end{acknowledgements}

\end{document}